\documentclass{PoS}

\def\G{\Gamma}

\def\s#1{{\scriptscriptstyle #1}}
\newcommand{\be}{\begin{equation}}
\newcommand{\bea}{\begin{eqnarray}}
\newcommand{\ee}{\end{equation}}
\newcommand{\eea}{\end{eqnarray}}

\def\gammaE{\gamma_\s{E}}

\def\2eqs#1#2{Eqs.~(\ref{#1}) and~(\ref{#2})}
\def\3eqs#1#2#3{Eqs.~(\ref{#1}),~(\ref{#2}) and~(\ref{#3})}

\def\operhs{\Upsilon}
\def\homo{\kappa}
\def\Y{\Xi}
\def\fg{\mathrm{I}\!\Gamma}

\def\diff{\mathrm{d}}

\title{Background field dependence from the Slavnov-Taylor identity in (non-perturbative) Yang-Mills theory}

\ShortTitle{Background field dependence by the ST identity}

\author{\speaker{Andrea Quadri}
       \\
        Universit\`a degli Studi di Milano and INFN, Sez. di Milano\\
	   via Celoria 16, I-20133 Milan, Italy\\
        E-mail: \email{andrea.quadri@mi.infn.it}}

\abstract{We show that in Yang-Mills theory 
the Slavnov-Taylor (ST) identity, extended in the presence of a background gauge connection,
allows to fix in a unique way the dependence of the vertex functional on the background, once the 1-PI
amplitudes at zero background are known. The reconstruction of the background dependence is carried out
by purely algebraic techniques and therefore can be applied in a non-perturbative scheme (e.g. on the lattice or in the
Schwinger-Dyson approach), provided that the latter preserves the ST identity.
The field-antifield redefinition, which replaces the classical background-quantum splitting when 
quantum corrections are taken into account, is considered on the example
 of an instanton background in SU(2) Yang-Mills theory.}

\FullConference{International Workshop on QCD Green's Functions, Confinement and Phenomenology,\\
		September 05-09, 2011\\
		Trento Italy}

\begin{document}

\section{Introduction}

In a recent paper \cite{Binosi:2011ar} the background field method
(BFM) \cite{DeWitt:1967ub,Abbott:1980hw} has been
reformulated within the Batalin-Vilkovisky formalism
\cite{Batalin:1977pb} as a prescription for handling the
quantization of a gauge theory in the presence of a
topologically non-trivial background  $\widehat{A}_\mu$.
The background-dependent amplitudes are recovered via a canonical transformation. 
The latter guarantees the fulfillment of the relevant Slavnov-Taylor
(ST) identity of the model at non-zero background.

Since the approach only relies on the ST identity associated
with the BRST symmetry of the theory (extended in the presence
of the background connection $\widehat{A}_\mu$), it can be applied
in any non-perturbative symmetric framework which preserves
the relevant functional identities of the theory, like e.g. the approach to QCD based
on the Schwinger-Dyson (SD) equations. 
Moreover our formalism provides a consistent strategy for the implementation of the
BFM on the lattice. The relevant composite operators, needed to control
the background dependence at the quantum level, are identified
in terms of the set of antifields of the theory plus an additional
anticommuting source $\Omega_\mu$, coupled to the covariant
derivative of the antighost field.  
Once this set of operators is introduced on the lattice, one can reconstruct
the background-dependent amplitudes in a unique way. 
We remark that, if one is only interested
to the gauge sector, this set of operators boils down to just two: the covariant derivative w.r.t.
the ghost, coupled to the antifield $A^*_\mu$ (i.e. the source of the
BRST transformation of the gauge field), and the already mentioned covariant derivative w.r.t. the antighost.

The implementation of the BFM on the lattice (for whatever value of the gauge fixing parameter)
would be a long awaited leap forward~\cite{Dashen:1980vm}. 
For instance, the understanding of the behaviour of the ghost and gluon propagators in the deep IR,
where the existence
of massive solutions has been  firmly established in the Landau gauge by
recent lattice data~\cite{Cucchieri:2007md,Bogolubsky:2007ud} supported by
SD-computations \cite{Aguilar:2006gr}-\cite{Binosi:2009qm},
might benefit from the extension
of these investigations to different gauges and to the effects
due to the presence of non-trivial backgrounds. 

The existence of a canonical transformation, implementing
at the quantum level the deformation of the background-quantum
splitting, has several important consequences.
On the one hand, it shows that the BFM can be made consistent with
the fundamental symmetries of the theory also in 
non-perturbative approaches to Yang-Mills theory.
On the other hand, it allows us to reconstruct the full
dependence of the vertex functional on the background
connection by purely algebraic means.

This in turn provides a separation between the integration
over the quantum fluctuations around the classical background
(controlled by the vertex functional $\fg_0$ at zero
background), and the further reconstruction of the background-dependent
amplitudes.
The latter can be represented in a compact way by using
homotopy techniques \cite{Binosi:2011ar}.

Another interesting consequence is that the background splitting at the quantum
level implies a (gauge background-dependent) 
redefinition also of the ghosts and the antifields, which, to the best
of our knowledge, was not pointed out before in the literature.

\medskip
The paper is organized as follows.
In Sect.~\ref{sect.1} we introduce our notation.
In Sect.~\ref{sect.2} we summarize the results
of the implementation of the BFM via a canonical
transformation, culminating in the homotopy formula
for the vertex functional.
In Sect.~\ref{sect.3} we show that the canonical
transformation, governing the BFM splitting, gives rise
to a field and antifield redefinition both in the gauge
and the ghost sector.
In Sect.~\ref{sect.4} we discuss as an example
the explicit deformation at one loop level of the instanton solution
in the singular gauge for SU(2) pure Yang-Mills theory.
Conclusions are finally presented in Sect.~\ref{sect.5}.

\section{Classical Action, BV Bracket and ST identity}
\label{sect.1}

We consider Yang-Mills theory based on a semisimple gauge
group $G$ with generators $T_a$ in the adjoint representation, satisfying
\begin{eqnarray}
[ T_a, T_b] = i f_{abc} T_c \, .
\label{e.1}
\end{eqnarray}
The Yang-Mills action $S_{YM}$ is 
\begin{eqnarray}
S_{YM} = -\frac{1}{4g^2} \int d^4x \,  G_{a\mu\nu}
G_a^{\mu\nu}
\label{e.2}
\end{eqnarray}
where $g$ is the coupling constant and $G_{a\mu\nu}$ is the
Yang-Mills field strength
\begin{eqnarray}
G_{a\mu\nu} = \partial_\mu A_{a\nu} - \partial_\nu A_{a\mu}
+ f_{abc} A_{b\mu} A_{c\nu} \, .
\label{e.3}
\end{eqnarray}
We adopt a (background) $R_\xi$-gauge-fixing condition \cite{QTS7} by
adding to $S_{YM}$ the gauge-fixing term
\begin{eqnarray}
S_{g.f.} & =& \int d^4x \, s \Big [ \bar c_a \Big (
\frac{\xi}{2} B_a - D_\mu[\hat A] (A-\hat A)_a \Big ) \Big ] 
\nonumber \\
& = & \int d^4x \, \Big ( \frac{\xi}{2} B_a^2
 - B_a D_\mu[\hat A] (A-\hat A)_a \nonumber \\
& & ~~~~~~~~ + \bar c_a D_\mu[\hat A] (D^\mu[A] c)_a
+ (D_\mu[A] \bar c)_a \Omega_a^\mu \Big ) \, .
\label{e.4}
\end{eqnarray}
In the above equation $\xi$ is the gauge parameter (the Landau
gauge used in \cite{Binosi:2011ar} is obtained for $\xi=0$) and $\hat A_{a\mu}$
denotes the background connection. 
$\bar c_a,c_a$ are the antighost and ghost fields respectively
and $B_a$ is the Nakanishi-Lautrup multiplier field.

We will sometimes use the notation $A_\mu = A_{a\mu} T_a$
and similarly for $\hat A_\mu, \bar c, c, B$.

The BRST differential $s$ acts on the fields of the theory as follows
\begin{eqnarray}
&& s A_{a\mu} = D_\mu[A] c_a \equiv \partial_\mu c_a + f_{abc} A_{b\mu}
 c_c \, , \nonumber \\
&& s c_a = -\frac{1}{2} f_{abc} c_b c_c \, , \nonumber \\
&& s \bar c_a = B_a \, , \quad s B_a =0 \, .
\label{e.5}
\end{eqnarray}
$s$ is nilpotent.
$\Omega_{a\mu}$ is an external source 
\cite{Grassi:1995wr}-\cite{Ferrari:2000yp} 
with ghost number $+1$
pairing with the background connection $\hat A_{a\mu}$ into a BRST
doublet \cite{Quadri:2002nh}
\begin{eqnarray}
s \hat A_{a\mu} = \Omega_{a\mu} \, , \qquad s \Omega_{a\mu} = 0 \, .
\label{e.6}
\end{eqnarray}

$\Omega_{a\mu}$ was introduced in \cite{Grassi:1995wr}, where it was
shown that no new  $\widehat A_\mu$- 
and $\Omega_\mu$-dependent anomalies 
can arise, as a consequence of the pairing in
eq.(\ref{e.6}).

Since the BRST transformations of the fields $A_{a\mu}$ and
$c_a$ in eq.(\ref{e.5}) are non-linear in the quantum fields,
we need a suitable set of sources,  known as antifields \cite{af,Gomis:1994he}, in order 
to control their quantum corrections.
For that purpose we finally add to the classical action the following 
antifield-dependent term
\begin{eqnarray}
S_{a.f.} = \int d^4x \, \Big ( A_{a\mu}^* D^\mu[A] c_a -
c_a^* \Big ( -\frac{1}{2} f_{abc} c_b c_c \Big ) - \bar c^*_a B_a\Big ) \, .
\label{e.6.1}
\end{eqnarray}
Although it is not necessary for renormalization purposes, we have included
in eq.(\ref{e.6.1}) the antifield $\bar c^*_a$ for $\bar c_a$.
This will allow us to treat on an equal footing all the fields of the
theory by a single Batalin-Vilkovisky (BV) bracket \cite{Batalin:1977pb,Gomis:1994he}.

We summarize in Table~\ref{tableI} the ghost charge,
statistics and dimension of the fields and antifields of the theory.
We  end up with the  tree-level vertex functional given by
\begin{eqnarray}
\fg^{(0)} & = & S_{YM} + S_{g.f.} + S_{a.f.} \, .
\label{e.6.2}
\end{eqnarray}
\begin{table}
\begin{center}
\begin{tabular}{r||c|c|c|c|c|c|c|c|c|c|}
 & $\ A_{a\mu}\ $ &  $\ c_a\ $ & $\ \bar c_a\ $  & $\ B_a\ $ &  $\ A^{*}_{a\mu} \ $ & $\ c^{*}_a\ $  
& $\ {\bar c}_a^* \ $ & $\ B_a^* \ $ 
& $\ \widehat{A}_{a\mu}\ $ & $\ \Omega_{a\mu}\ $\\
\hline\hline
Ghost charge & 0  & 1 & -1  & 0  & -1  & -2 & 0 & -1 & 0 & 1\\
\hline
Statistics  & B & F & F  & B & F &  B & B & F & B & F\\
\hline
Dimension & 1 &  0 & 2 & 2 & 3 &  4  & 2  & 2 & 1 & 1 \\
\hline
\end{tabular} 
\caption{Ghost charge, statistics (B for Bose, F for Fermi), and mass dimension of both the Yang-Mills conventional fields and anti-fields as well as background fields and sources. \label{tableI}}
\end{center}
\end{table}

\medskip

$\fg^{(0)}$ fulfills several functional identities \cite{Binosi:2011ar}:
\begin{itemize}
\item the Slavnov-Taylor (ST) identity
\medskip

The ST identity encodes in functional form the invariance under the
BRST differential $s$ in eqs.(\ref{e.5}) and (\ref{e.6}). In order to set up the formalism required
for the consistent treatment of the quantum deformation for the
background-quantum splitting, it is convenient to write the
ST identity within the BV formalism.

We adopt for the BV bracket the same conventions as in~\cite{Gomis:1994he}; then, using only left derivatives, one can write
\begin{eqnarray}
(X,Y) = \int\!\diff^4x \sum_\phi
\left[ (-1)^{\epsilon_{\phi} (\epsilon_X+1)}
\frac{\delta X}{\delta \phi} \frac{\delta Y}{\delta \phi^*}
- (-1)^{\epsilon_{\phi^*} (\epsilon_X+1)}
\frac{\delta X}{\delta \phi^*} \frac{\delta Y}{\delta \phi}
\right]
\label{bracket}
\end{eqnarray}
where the sum runs over the fields $\phi = \{A_{a\mu},c_a,\bar c_a,B_a \}$ and the antifields 
$\phi^* = \{ A^*_{a\mu}, c^*_a, \bar c_a^*, B^*_a \}$. In the equations above, 
$\epsilon_\phi$, $\epsilon_{\phi^*}$ and $\epsilon_X$ represent respectively the grading of the field $\phi$, the antifield $\phi^*$ and the functional $X$.

The extended ST identity arising from the invariance
of $\fg^{(0)}$ under the BRST differential
in eq.(\ref{e.5}) and eq.(\ref{e.6}), in the presence of a background field,
can now be written as
\begin{eqnarray}
\int\!\diff^4x\, \Omega_a^\mu(x)
\frac{\delta \fg^{(0)}}{\delta \widehat A^a_\mu(x)} = 
- \frac{1}{2}\, (\fg^{(0)},\fg^{(0)}) .
\label{m.1}
\end{eqnarray}
\item the B-equation
\begin{eqnarray}
\frac{\delta \fg^{(0)}}{\delta B_a} = \xi B_a - D_\mu[\hat A] (A-\hat A)_a - \bar c^*_a\, .
\label{b.eq}
\end{eqnarray}
The B-equation  guarantees the stability of the gauge-fixing condition under radiative corrections.
Notice that the r.h.s. of the above equation is linear in the quantum fields
and thus no new external source is needed in order to define it.
It does not receive any quantum corrections.
\item the antighost equation
\begin{eqnarray}
\frac{\delta \fg^{(0)}}{\delta \bar c_a} =  D[\hat A]_\mu
\frac{\delta \fg^{(0)}}{\delta A_{a\mu}^*} -
D_\mu[A] \Omega_{a\mu} \, .
\label{antigh.eq}
\end{eqnarray}
In the background Landau gauge one can also write an equation
for the derivative of the effective action w.r.t. the ghost
$c_a$ (also sometimes called antighost equation) \cite{hep-th/0405104}. 
This was introduced in \cite{hep-th/9804013} in the context of the 
BFM formulation of Yang-Mills theory for semi-simple gauge groups in the background 't Hooft gauge.

\item the background Ward identity
\medskip

By using the background gauge-fixing condition in eq.(\ref{e.4}),
the vertex functional $\fg^{(0)}$ becomes invariant under
a simultaneous gauge transformation of the quantum fields,
external sources and the background connection, i.e.
\begin{eqnarray}
\!\!\!\!\!\!\!\!\!\!\!\!\!\!\!\!\!\!
{\cal W}_a \fg^{(0)} & = & - \partial_\mu \frac{\delta \fg^{(0)}}{\delta 
\hat A_{a\mu}} + f_{acb} \hat A_{b\mu} \frac{\delta \fg^{(0)}}{\delta \hat A_{c\mu}} - \partial_\mu \frac{\delta \fg^{(0)}}{\delta 
\hat A_{a\mu}} + f_{acb} A_{b\mu} \frac{\delta \fg^{(0)}}{\delta A_{c\mu}} + \sum_{\Phi \in \{ B,c,\bar c \}} f_{acb} \Phi_b \frac{\delta \fg^{(0)}}{\delta \Phi_c} \nonumber \\
\!\!\!\!\!\!\!\!\!
& & 
      + f_{acb} A^*_{b\mu} \frac{\delta \fg^{(0)}}{\delta A^*_{c\mu}}
      + f_{acb} c^*_b \frac{\delta \fg^{(0)}}{\delta c^*_c} 
      + f_{acb} {\bar c}^*_b \frac{\delta \fg^{(0)}}{\delta \bar{c^*}_c}= 0 \, .
\label{w.id}
\end{eqnarray}

\end{itemize}

Several comments are in order here. First we remark
that the ST identity (\ref{m.1}) is bilinear in the vertex functional,
unlike the background Ward identity (\ref{w.id}).
Thus the relations between 1-PI amplitudes, derived by functional differentiation  of the
ST identity in eq.(\ref{m.1}),
are bilinear, in contrast with the linear ones generated 
by functional differentiation of the background Ward identity
(\ref{w.id}). 
The linearity of the background Ward identity explains why the BFM
has been advantageously used in several applications, ranging from 
perturbative calculations in Yang-Mills theories~\cite{Abbott:1980hw,Ichinose:1981uw} and in the
 Standard Model~\cite{Denner:1994xt,hep-ph/0102005} to gravity and supergravity calculations~\cite{Gates:1983nr}.

One should however notice that the background Ward identity
is no substitute to the ST identity: physical unitarity
stems from the validity of the ST identity and does not follow from the background Ward identity alone 
\cite{Ferrari:2000yp}.

\noindent
Since the theory is non-anomalous, in perturbation theory 
all the
functional identities in eqs.~(\ref{m.1}), (\ref{b.eq}), (\ref{antigh.eq})
and (\ref{w.id}) are fulfilled also for the full
vertex functional $\fg$ \cite{hep-th/9807191,hep-th/9905192}. This can be proven in a
regularization-independent way by standard
methods in Algebraic Renormalization \cite{Becchi:1999ir,Ferrari:2000yp,hep-ph/9907426}.
In what follows
we assume that the same identities hold true for the 
vertex functional of the theory in the non-perturbative
regime.

\section{Canonical Transformation for the Background Dependence}\label{sect.2}

In order to control the dependence on the background connection
we start from eq.(\ref{m.1})  for the full vertex functional $\fg$:
\begin{eqnarray}
\int d^4x \, \Omega_{a\mu}(x) \frac{\delta \fg}{\delta \hat A_{a\mu}(x)} =
-\frac{1}{2} (\fg,\fg) \, .
\label{c.1}
\end{eqnarray}
By taking a derivative w.r.t. $\Omega_{a\mu}(x)$ and then setting
$\Omega_{a\mu}=0$ we get
\begin{eqnarray}
\left . \frac{\delta \fg}{\delta \hat A_{a\mu}(x)} \right |_{\Omega_\mu=0} = -
 ( \left . \frac{\delta \fg}{\delta \Omega_{a\mu}(x)} \right |_{\Omega_\mu=0} ,\left . \fg \right |_{\Omega_\mu=0}) \, .
\label{c.2}
\end{eqnarray}
This equation states that the derivative of the full vertex functional
$\fg$  w.r.t. $\hat A_{a \mu}$ at $\Omega_\mu=0$ equals the variation of 
$\fg$ w.r.t. to a canonical transformation \cite{Gomis:1994he} generated by the
fermionic functional 
$ \left . \frac{\delta \fg}{\delta \Omega_{a\mu}(x)} \right |_{\Omega_\mu=0}$.

This a crucial observation. First of all
 it shows that the source
$\Omega_{a\mu}$ has a clear geometrical interpretation,
being  the source of the fermionic functional which governs
the canonical transformation  giving rise to the background field dependence.
Moreover, the dependence of the vertex functional on the
background field is designed in such a way to preserve the
validity of the ST identity (since the transformation is canonical).

In a non-perturbative setting, we can use eq.(\ref{c.2}) in order to
control the background-dependent amplitudes.
For that purpose one needs a method for solving
eq.(\ref{c.2}). An effective recursive procedure is based
on cohomological techniques.
Let us introduce the auxiliary BRST differential $\omega$
given by \cite{Binosi:2011ar}
\begin{eqnarray}
\omega \hat A_{a \mu} = \Omega_{a\mu} \, , ~~~~
\omega \Omega_{a\mu} = 0 \, ,
\label{c.5}
\end{eqnarray}
while $\omega$ does not act on the other variables of the theory.
Clearly $\omega^2=0$ and, since the pair $(\hat A_{a\mu}, \Omega_{a\mu})$ forms a BRST doublet \cite{Quadri:2002nh} under $\omega$,
the cohomology of $\omega$ in the space of local functionals
spanned by $\hat A_{a\mu}, \Omega_{a\mu}$ is trivial.

This allows us to introduce the homotopy operator $\kappa$
according to
\begin{eqnarray}
\kappa = \int d^4x \, \int_0^1 dt \, \hat A_{a\mu}(x) \lambda_t \frac{\delta} {\delta \Omega_{a\mu}(x)}
\label{c.6}
\end{eqnarray}
where the operator $\lambda_t$ acts as follows on a functional
$X(\hat A_{a\mu}, \Omega_{a\mu}; \zeta)$
\begin{eqnarray}
\lambda_t X(\hat A_{a\mu}, \Omega_{a\mu}; \zeta) = 
X(t \hat A_{a\mu}, t \Omega_{a\mu}; \zeta)
\label{c.7}
\end{eqnarray}
depending
on $\hat A_{a\mu}, \Omega_{a\mu}$ and on other variables
collectively denoted by $\zeta$.
The operator $\kappa$ obeys the relation
\begin{eqnarray}
\{ \omega, \kappa \} = {\bf 1}_{\hat A,\Omega}
\label{c.8}
\end{eqnarray}
where ${\bf 1}_{\hat A,\Omega}$ denotes the identity
in the space of functionals containing at least one $\hat A_\mu$ or $\Omega_\mu$.

Then we can rewrite the ST identity (\ref{c.1}) as
\begin{eqnarray}
\omega \fg = \Upsilon \, ,
\label{c.111}
\end{eqnarray}
where
\begin{eqnarray}
\Upsilon = - \frac{1}{2}(\fg,\fg) \, .
\label{c.112}
\end{eqnarray}
By the nilpotency of $\omega$ 
\begin{eqnarray}
\omega \Upsilon = 0 \, .
\label{c.112.1}
\end{eqnarray}
Since $\left . \Upsilon \right |_{\Omega=0}=0$,
we have from eq.(\ref{c.8})
\begin{eqnarray}
\Upsilon = \{ \omega, \kappa \} \Upsilon = \omega \kappa \Upsilon
\label{c.113}
\end{eqnarray}
Thus from eq.(\ref{c.111}) we have the identity 
\begin{eqnarray}
\omega ( \fg - \kappa \Upsilon) = 0 \, ,
\label{c.114}
\end{eqnarray}
which has the general solution
\begin{eqnarray}
\fg = \fg_0 + \omega \Xi +\kappa \Upsilon
\label{c.10}
\end{eqnarray}
with $\Xi$ an arbitrary functional with ghost number $-1$.
In the above equation $\fg_0$ denotes the vertex functional
evaluated at $\hat A_\mu = \Omega_\mu =0$ (i.e.
the set of 1-PI amplitudes with no background insertions and
no $\Omega_\mu$-legs). The second term vanishes
at $\Omega_\mu=0$ but is otherwise unconstrained. 
I.e. the extended ST identity is unable to fix
 the sector where $\Omega_\mu \neq 0$.
However this ambiguity is irrelevant if one is interested
in the 1-PI amplitudes with no $\Omega_\mu$-legs, which
are those needed for physical computations.

In practical applications 
it is convenient to expand the  term $\kappa \Upsilon$ in eq.(\ref{c.10})
according to the number
of background legs. Then one can write a tower of equations
allowing to solve for the dependence on $\hat A_{a \mu}$
recursively down to the 
vertex functional at zero background $\fg_0$ \cite{paperII}.

In the zero background ghost sector $\Omega_\mu=0$, the $\omega \Y$ term in Eq.~(\ref{c.10}) drops out, and one is left with the result 
(notice that $\fg_{\bar c^*_a} = - B_a$)
\begin{eqnarray}
\!\!\!\!\!\!\!\!\!\!\!
\left . \fg \right |_{\Omega=0}&=&\homo\operhs+\fg_0\nonumber \\
&=& - \left . \int\! \mathrm{d}^4x \,{\widehat{A}^a_\mu(x)}\!\int_0^1\!\!\mathrm{d}t\,\lambda_t\,\frac{\delta}{\delta_{\Omega^\mu_a(x)}}\!\int\!\mathrm{d}^4y
\left[\fg_{A^{*\nu}_b}(y)\fg_{A^{b}_\nu}(y) \right . \right . \nonumber \\
&& \qquad  \left . \left . +\fg_{c^{*b}}(y)\fg_{c^b}(y)+B^b(y)\fg_{\bar c^b}(y)\right]\right|_{\Omega_\mu=0}
\nonumber \\
&&+ \fg_0.
\label{int.rep.g}
\end{eqnarray}
In the above equation we have used the short-hand notation 
$\fg_\varphi = \frac{\delta \fg}{\delta \varphi}$.
%


\section{Field and Antifield Redefinition in the BFM}\label{sect.3}

By taking a derivative w.r.t. $\Omega_{a\mu}$ of eq.(\ref{m.1}) and then setting $\Omega_\mu=0$
we get (from now on we denote by $\G$ the vertex functional where $\Omega_\mu$ is set to zero)
\bea
&& 
\!\!\!\!\!\!\!\!\!\!\!\!\!\!\!\!\!
\frac{\delta \G}{\delta \widehat A_{a\mu}} = - \int d^4x \, \Big (
\left . \frac{\delta^2 \fg}{\delta \Omega_{a\mu} \delta A^*_{b\nu}} \right |_{\Omega_\mu=0} 
         \frac{\delta \G}{\delta A_{b\nu}} -
\frac{\delta \G}{\delta A^*_{b\nu}} \left . \frac{\delta^2 \fg}{\delta \Omega_{a\mu} \delta A_{b\nu}} \right |_{\Omega_\mu = 0}
\nonumber \\
&& \qquad \qquad \quad
-\left . \frac{\delta^2 \fg}{\delta \Omega_{a\mu} \delta c^*_{b}} \right |_{\Omega_\mu=0} 
         \frac{\delta \G}{\delta c_{b}} -
\frac{\delta \G}{\delta c^*_{b}} \left . \frac{\delta^2 \fg}{\delta \Omega_{a\mu} \delta c_{b}} \right |_{\Omega_\mu = 0}
\nonumber \\
&& \qquad \qquad \quad
-\left . \frac{\delta^2 \fg}{\delta \Omega_{a\mu} \delta {\bar c}^*_{b}} \right |_{\Omega_\mu=0} 
         \frac{\delta \G}{\delta {\bar c}_{b}} -
\frac{\delta \G}{\delta {\bar c}^*_{b}} \left . \frac{\delta^2 \fg}{\delta \Omega_{a\mu} \delta {\bar c}_{b}} \right |_{\Omega_\mu = 0}
\Big ) \, .
\label{redef.1}
\eea
Suppose  that one can find a set of field and antifield redefinitions
\bea
& A_{a\nu} \rightarrow A_{a\nu} - {\cal G}_{a\nu} \, , & A^*_{a\nu} \rightarrow A^*_{a\nu} - {\cal G}^*_{a\nu} \, , 
\nonumber \\
& c_a \rightarrow c_a + {\cal C}_a \, , &  c_a^* \rightarrow c_a^* + {\cal C}_a^* \, , \nonumber \\
& \bar c_a \rightarrow \bar c_a + \bar{\cal C}_a \, , & \bar c_a^* \rightarrow \bar c_a^* + \bar {\cal C}_a^* \, ,  
\label{redef.2}
\eea
such that
\bea
&&
\frac{\delta {\cal G}_{b\nu}}{\delta \widehat A_{a\mu}} = \left . \frac{\delta^2 \fg}{\delta \Omega_{a\mu} \delta A^*_{b\nu}} \right |_{\Omega_\mu=0}  \, , \qquad
\frac{\delta {\cal G}^*_{b\nu}}{\delta \widehat A_{a\mu}} = \left . \frac{\delta^2 \fg}{\delta \Omega_{a\mu} \delta A_{b\nu}} \right |_{\Omega_\mu=0}  \, ,  \nonumber \\
&&
\frac{\delta {\cal C}_{b}}{\delta \widehat A_{a\mu}} = \left . \frac{\delta^2 \fg}{\delta \Omega_{a\mu} \delta c^*_{b}} \right |_{\Omega_\mu=0}  \, , \qquad ~~
\frac{\delta {\cal C}^*_{b}}{\delta \widehat A_{a\mu}} = \left . \frac{\delta^2 \fg}{\delta \Omega_{a\mu} \delta  c_{b}} \right |_{\Omega_\mu=0}  \, ,  \nonumber \\
&&
\frac{\delta \bar {\cal C}_{b}}{\delta \widehat A_{a\mu}} = \left . \frac{\delta^2 \fg}{\delta \Omega_{a\mu} \delta {\bar c}^*_{b}} \right |_{\Omega_\mu=0}  \, , \qquad ~~
\frac{\delta \bar {\cal C}^*_{b}}{\delta \widehat A_{a\mu}} = \left . \frac{\delta^2 \fg}{\delta \Omega_{a\mu} \delta  {\bar c}_{b}} \right |_{\Omega_\mu=0}  \, .  \nonumber \\
\label{redef.3}
\eea
Then the solution to eq.(\ref{redef.1}) is obtained by carrying out the field and antifield redefinition in eq.(\ref{redef.2}) on the vertex
funtional at zero background $\G[A_\mu, c, \bar c, A^*_\mu, c^*, \bar c^*;0]$ according to
\be
\G[A_\mu, c, \bar c, A^*_\mu, c^*, \bar c^*; \widehat A_\mu] = 
\G[A_\mu - {\cal G}_\mu, c + {\cal C}, \bar c + \bar {\cal C}_a, A^*_\mu - {\cal G}^*_\mu, c^* + {\cal C}^*, \bar c^* + \bar{\cal C}^*; 0] \, .
 \ee

The background-dependent field and antifield redefinition in eq.(\ref{redef.2})
generalizes the classical background-quantum splitting and is the correct mapping
when quantum corrections are taken into
account. This result directly follows from the requirement of the validity of the ST identity.
We remark that the redefinition in eq.(\ref{redef.2}) also involves the ghosts and the antifields.. This is in sharp
contrast with the classical background-quantum splitting, which
is limited to the gauge field.

The existence of the field and antifield redefinitions in eq.(\ref{redef.2}) requires a careful check of the corresponding
integrability conditions. This has been done for the case of the gauge field in Ref.~\cite{Binosi:2011ar} and requires
an extensive use of the relations among 1-PI amplitudes encoded in the ST identity.
The analysis of the general case will be deferred to a later work. Here we only wish to remark that the
field and antifield redefinitions are related to the deformation of the canonical variables controlled by the canonical transformation generated by 
$\left . \frac{\delta\fg}{\delta \Omega_\mu} \right |_{\Omega_\mu = 0}$.

\section{One-loop Deformed Instanton Profile}\label{sect.4}

As an example, we sketch the one-loop corrections to the classical instanton \cite{'tHooft:1976fv} profile function. 
For a detailed treatment we refer the reader to \cite{paperII}.
To lowest order, the background-dependent field redefinition for $A_\mu$ 
in the first of Eqs.~(\ref{redef.2}) 
 implies that the background field will be deformed according to
\be
V^a_\mu(x)=\widehat{A}^a_\mu(x)+\int\!\diff^4y\,\Gamma_{\Omega^a_\mu A^{*b}_\nu}(y,x)\widehat{A}^b_\nu(y),
\ee
where the auxiliary function $\Gamma_{\Omega A^*}$ is evaluated at $\widehat{A}=0$; equivalently, in momentum space one has
\be
V^a_\mu(p)=\left[g_{\mu\nu}\delta^{ab}+\Gamma_{\Omega^a_\mu A^{*b}_\nu}(p)\right]\widehat{A}^b_\nu(p).
\ee
Notice that the formulas above are totally general and not limited to the instanton case we are considering here; thus the calculation of $\Gamma^{(1)}_{\Omega^a_\mu A^{*b}_\nu}$ performed below will determine the universal (lowest-order) deformation of any background
at one loop level.


The function $\Gamma_{\Omega^a_\mu A^{*b}_\nu}$ can be 
decomposed 
%
according to the following form factors:
\be
\Gamma_{\Omega^a_\mu A^{*b}_\nu}(p)=-\delta^{ab}\frac{g^2C_A}{16\pi^2}\left[A(p)g_{\mu\nu}+B(p)\frac{p_\mu p_\nu}{p^2}\right],
\ee
we see that in the instanton case the $B$ form factor does not contribute, and we finally get the one-loop corrected background field
\be
V^a_\mu(p)=\overline{\eta}^a_{\mu\nu}p_\nu \left[f_0(p)+f_1(p)\right];\qquad f_1(p)=-\frac{g^2C_A}{16\pi^2}A^{(1)}(p)f_0(p),
\label{1l-mom}
\ee
and $f_0$ is  the classical instanton profile
\bea
f_0(p)&=&\left(-8\pi^2\mathrm{i}\rho\right)\frac1{p^3}\left[-\frac2{p\rho}+K_1(p\rho)-(p\rho) K_1'(p\rho)\right]\nonumber \\
&=&\left(-8\pi^2\mathrm{i}\rho\right)\frac1{p^3}\left[-\frac2{p\rho}+(p\rho)K_2(p\rho)\right],
\label{effe}
\eea
with $K_i$ the modified Bessel functions of the second kind.

Choosing the Landau gauge (which is the appropriate choice in the instanton case) one has
at the one-loop level 
\be
\Gamma^{(1)}_{\Omega^a_\mu A^{*b}_\nu}(p)=-g^2C_A\delta^{ab}\int_k\frac1{k^2(k+p)^2}P_{\mu\nu}(k),
\ee
where $C_A$ is the Casimir eigenvalue of the adjoint representation
[$C_A=N$ for $SU(N)$]; a straightforward calculation gives (Euclidean space)
\bea
A^{(1)}(p)&=&-\frac32\frac1{d-4}+\frac32-\frac34\log\left(\frac{p^2}{\mu^2}\right)
\nonumber\\
B^{(1)}(p)&=&-\frac12.
\label{g.astar.omega}
\eea
The divergence in $A^{(1)}$ in the $d\to4$ limit is removed by adding a counterterm controlled
by the invariant ${\cal S}_0 (A^*_{a\mu} \hat A_{a\mu})$, where ${\cal S}_0$ is the linearized
ST operator ${\cal S}_0 = (\fg^{(0)}, \cdot )$.  

Eq.(\ref{g.astar.omega}) does not change in the one-loop approximation
if fermions are added to the theory. 

It is now convenient to have a representation of the instanton profile in position space; therefore we need to find the inverse Fourier transform of $f_1$. Let us set
\be
V^a_\mu(x)=\overline{\eta}^a_{\mu\nu}x_\nu\left[f_0(x) + f_1(x)\right];\qquad f_1(x)=\frac{\mathrm{i}}{4\pi^2}\frac{x_\nu}{r^2}\frac{\partial}{\partial x_\nu}\int_0^\infty\!\diff p\,p^3f_1(p)\frac1{pr}J_1(pr).
\ee
The evaluation of $f_1(x)$ can be performed analytically, and we find\footnote{This is only true in the singular gauge. In the regular gauge the integral over $p$ does not converge.}
\bea
f_1(x)&=&-3\frac{g^2C_A}{16\pi^2}\left[\frac1{\rho^2}\frac{1+\log\rho\mu}{\lambda^2(1+\lambda^2)}
-\frac{x_\nu}{r^2}\frac{\partial}{\partial x_\nu}\int_0^\infty\!\diff t\,  F(t,\lambda)\right], 
\eea
where we have  $t=p\rho$ and
\be
F(t,\lambda)=\log t\left[-\frac2t+tK_2(t)\right]\frac1{\lambda t}J_1(\lambda t).
\ee
The integral in $t$ yields
\bea
\int_0^\infty\!\diff t\,  F(t,\lambda) &=& \frac{1}{8\lambda^2}
\left\{ \log^2(1+\lambda^2) \lambda^2 -
4 \left( \log \frac{\lambda}{4} +2 \gammaE - 1 \right)
\lambda^2 \log \lambda 
+2 \lambda^2  {\rm Li}_2 \left( \frac{1}{1+\lambda^2} \right)\right.\nonumber\\
&+&\left. \left[  -2\lambda^2 \log \frac{\lambda^2}{1+\lambda^2}  +
\left(-2 + 4 \gammaE - 4 \log 2\right) \lambda^2 - 2\right] 
\log \left(1+\lambda^2\right) 
\right\},
\label{math}
\eea
where $\gammaE$ is the Euler-Mascheroni constant ($\gammaE=0.57721\dots$); thus one has
\be
\frac{x_\nu}{r^2}\frac{\partial}{\partial x_\nu}\int_0^\infty\!\diff t\,  F(t,\lambda) =\frac1{\rho^2}\left[-\frac{\gammaE-\log2}{\lambda^2(1+\lambda^2)}-\frac{\log\lambda}{\lambda^2}+\frac{1+\lambda^4}{2\lambda^4(1+\lambda^2)}\log(1+\lambda^2)\right],
\ee
which gives for $f_1$ the final result
\bea
f_1(\lambda)&=&-3\frac{g^2C_A}{16\pi^2}\frac1{\rho^2}\left[\frac{1+\log\rho\mu}{\lambda^2(1+\lambda^2)}+\frac{\gammaE-\log2}{\lambda^2(1+\lambda^2)}+\frac{\log\lambda}{\lambda^2}-\frac{1+\lambda^4}{2\lambda^4(1+\lambda^2)}\log(1+\lambda^2)\right]\!.\hspace{.8cm}
\eea

$f_1$ shows a log enhancement w.r.t. the classical profile both for $\lambda \rightarrow 0$ and
for $\lambda \rightarrow \infty$, i.e. both for small and large instanton sizes.
Clearly the one-loop corrected instanton is neither self-dual nor it reduces
to pure gauge as $r \rightarrow \infty$.

\section{Conclusions}\label{sect.5}

We have shown that there is a close connection between the quantization of Yang-Mills theory
in a topologically non-trivial background and the ST identity  of the theory (extended in the presence
of a background connection).

If the ST identity is fulfilled, the dependence of the vertex functional on the background
can be uniquely reconstructed (in the relevant sector at $\Omega_\mu=0$) by
algebraic techniques, starting from 1-PI amplitudes evaluated at zero background.

The procedure amounts to perform a field and antifield redefinition, controlled by a canonical
transformation w.r.t. the BV bracket associated with the ST identity. 
Moreover, a compact homotopy formula for the full vertex functional at non-zero background has been derived.

As an example of this technique, we have explicitly worked out in lowest order in the background field
the one-loop redefinition of the SU(2) instanton profile, induced by the canonical trasnformation
responsible for the quantum deformation of the classical background-quantum splitting.

These results could be applied to a variety of problems,  like e.g. the implementation of the BFM
on the lattice or SD-computations in the presence of a topologically non-trivial background.

Several aspects should be further investigated. We only mention a few of them here.
The fulfillment of the integrability conditions for eqs.(\ref{redef.3}) has to be further analyzed.
Although it is plausible that these conditions are indeed fulfilled (since they are a consequence
of the existence of a canonical transformation governing the dependence on the background field),
it would be very useful to obtain a more explicit form for the field and antifield
redefinition.

The SD equations for the $\Omega$-dependent kernels in eqs.(\ref{redef.3}) should be studied.
Finally one could also investigate whether the present approach, derived for the 1-PI vertex
functional $\fg$, can be extended to the well-known 2-PI formalism of \cite{Cornwall:1974vz}.

\section*{Acknowledgments}

Financial support from INFN and ECT$^*$ is gratefully acknowledged.

\end{document}